# Co-co-doping effect for (Ca,*RE*)FeAs$_2$ sintered bulk

H Yakita[1], H Ogino[1], A Sala[1,2,3], T Okada[1], A Yamamoto[1], K Kishio[1], A Iyo[2], H Eisaki[2] and J Shimoyama[1]

[1] Department of Applied Chemistry, The University of Tokyo, 7-3-1 Hongo, Bunkyo, Tokyo 113 8656, Japan

[2] National Institute of Advanced Industrial Science and Technology (AIST), Tsukuba, Ibaraki 05-8565, Japan

[3] University of Genova and CNR-SPIN, 16146 Genova, Italy

E-mail: 8757570603@mail.ecc.u-tokyo.ac.jp

Abstract

Superconducting properties of Co-co-doped (Ca,*RE*)FeAs$_2$ ((Ca,*RE*)112: *RE* = La, Pr) were investigated. Co-co-doping increased $T_c$ of (Ca,Pr)112 while Mn-co-doping suppressed superconductivity of (Ca,*RE*)112. Co-co-doped (Ca,La)112 showed large diamagnetic screening and sharper superconducting transition than Co-free (Ca,La)112. $T_c^{zero}$ observed in resistivity measurements increased from 14 K to 30 K by Co-co-doping, while $T_c^{onset}$ was not increased. The critical current density ($J_c$) of Co-co-doped (Ca,La)112 were approximately 2.1 x 10$^4$ Acm$^{-2}$ and 3.2 x 10$^3$ Acm$^{-2}$ at 2 K and 25 K, respectively, near zero field. These relatively high $J_c$s and large diamagnetic screening observed in susceptibility measurement as for polycrystalline bulks suggest bulk superconductivity of Co-co-doped (Ca,*RE*)112 compounds.

# 1. Introduction

Since the report of superconductivity in LaFeAs(O,F) with $T_c$ = 26 K[1], iron based superconductors have gathered great attention, resulting discoveries of several new iron-based superconductor families. Searches for new superconductors with anti-PbO type FeAs layer have been continued after the discovery of representative iron based superconductors(11[2], 111[3], 122[4], 1111[1]), and, furthermore, those with complex blocking layers, such as perovskite type oxide (*ex.* $Fe_2P_2Sr_4Sc_2O_6$[5]), PtAs (*ex.* $Ca_{10}(Pt_4As_8)(Fe_2As_2)_5$[6]), $BaTi_2As_2O$ (*ex.* $Ba_2Ti_2Fe_2As_4O$[7]), and $RE_4TeO_4$ layers (*ex.* $Pr_4Fe_2As_2TeO_4$[8]), have been successively found.

Recently new compounds with a new crystal structure (Ca,*RE*)FeAs$_2$ (called (Ca,*RE*)112) have been reported for *RE* = La-Gd[9-12]. (Ca,*RE*)112 are composed of stacking of Fe$_2$As$_2$ layers and (Ca,*RE*)$_2$As$_2$ blocking layers where As$_2$ form zig-zag chain between two (Ca,*RE*) planes[9,10]. The formal valence of As in the As$_2$ chain layers is considered to be -1, while that in Fe$_2$As$_2$ layers is -3. (Ca,*RE*)112 compounds have monoclinic crystal structure with space group $P2_1$[9] or $P2_1/m$[10]. In our previous study, the high pressure synthesis was found to be an effective way to synthesize (Ca,*RE*)112 compounds with relatively small *RE* ions, such as Sm, Eu and Gd. We have also reported the relationship between $T_c$ and distance between iron layers $d_{Fe-Fe}$ of (Ca,*RE*)112 synthesized with nominal compositions of $(Ca_{0.85}RE_{0.15})FeAs_2$. (Ca,*RE*)112 with *RE*=La, Pr, Nd, Sm, Eu and Gd showed superconductivity, while (Ca,Ce)112 did not show superconductivity down to 2 K. Kudo *et al.* reported that Sb-co-doping to (Ca,*RE*)112 increased $T_c$ to 47, 43, 43, 43 K for *RE* = La, Ce, Pr, Nd, respectively[13]. Sb-co-doping increases *b*-axis length accompanying improvement of As-Fe-As bond angle. This report revealed that isovalent doping can enhance $T_c$ of (Ca,*RE*)112, while electron or hole doping effect have not been studied.

Recently, $J_c$ and anisotropy of $(Ca_{0.77}La_{0.18})Fe_{0.9}As_2$ single crystals have been evaluated[14]. The $J_c(H//c)$ of the $(Ca_{0.77}La_{0.18})Fe_{0.9}As_2$ single crystal exceeded $10^5$ Acm$^{-2}$ at 5 K under the magnetic fields below 4 T. The anisotropy factor $\gamma$ (= $H_{c2}(//ab)$ / $H_{c2}(//c)$) was approximately 2.0-4.2 in $(Ca_{0.77}La_{0.18})Fe_{0.9}As_2$, which is smaller than that of 1111 type superconductors.

In this paper, we report effects of transition metals (*TM*: Co, Mn) co-doping on superconducting properties of (Ca,La)112 and (Ca,Pr)112. Effects of additional electron or hole doping on superconducting properties of this system were investigated. Improvement of superconducting properties with $T_c$ up to 38 K were observed in Co-co-doped (Ca,*RE*)112 in spite of direct impurity doping to Fe site contrary to the case of 122 and 1111 type superconductors[15, 16].

## 2. Experimental details

All samples were synthesized by the solid state reaction, starting from FeAs(3N), LaAs(3N), PrAs(3N), Ca(2N), As(4N), MnAs(3N), and CoAs(3N) powders. These powders were mixed and pelletized in an argon-filled glove box. Heat treatment was carried out at 1000°C for 24 h in alumina crucibles which were sealed in evacuated quartz ampoules prior to heating, or at 1050°C for 1 h under a pressure of 2 GPa in a boron nitride (BN) cell using a wedge-type cubic-anvil high-pressure apparatus (Riken CAP-07). The constituent phases were investigated by powder XRD measurements (RIGAKU Ultima-IV) using Cu K$_\alpha$ radiation, and the intensity data were collected in the 2$\theta$ range of 5-80° in increments of 0.02°. Microstructure observation and compositional analysis were performed using a scanning electron microscopy (SEM; Hitachi High-Technologies TM3000) equipped with energy dispersive X-ray spectrometry (EDX; Oxford Instruments SwiftED 3000). The magnetic susceptibility was measured by a SQUID magnetometer (Quantum Design MPMS-XL5s). The electrical resistivity was measured by the AC four-point-probe method (Quantum Design, Physical Property Measurement System).

## 3. Results and discussion

All synthesized bulk samples were composed of plate-like (Ca,*RE*)112 crystals and impurity phases. Typical images of obtained plate-like crystals were shown in Fig.1. Figure 2 shows powder XRD patterns of polycrystalline samples of (Ca$_{0.9}$*RE*$_{0.1}$)(Fe$_{0.97}$Mn$_{0.03}$)As$_2$ sintered at 1000°C for 24 h in evacuated quartz ampoules, and (Ca$_{0.9}$*RE*$_{0.1}$)FeAs$_2$ and (Ca$_{0.9}$*RE*$_{0.1}$)(Fe$_{1-x}$Co$_x$)As$_2$ ($x$ = 0.03, 0.05) sintered at 1050°C for 1 h under 2 GPa (*RE* = La, Pr). (Ca,*RE*)112 phase formed in the all samples and FeAs and FeAs$_2$ phases were detected as impurity phases. Diffraction peaks of BN detected in *TM*-free and Co-co-doped samples is originated in high pressure media. XRD patterns coincide with the growth of plate-like grains in polycrystalline samples, because intensity of the 002 peak (~17°) is higher or almost same compared to that of main peak (~37°), while simulated intensity of 002 peak is approximately 20% of that of the main peak. $d_{\text{Fe-Fe}}$ values calculated from the XRD patterns are decreased 0.01-0.02 Å compared to *TM*-free samples by Co-co-doping. This result is similar to the decrease of *c*-axis length by Co doping in LaFeAsO[17] and CaFeAsH[18], indicating that Co-co-doping also works as electron doping in (Ca,*RE*)112.

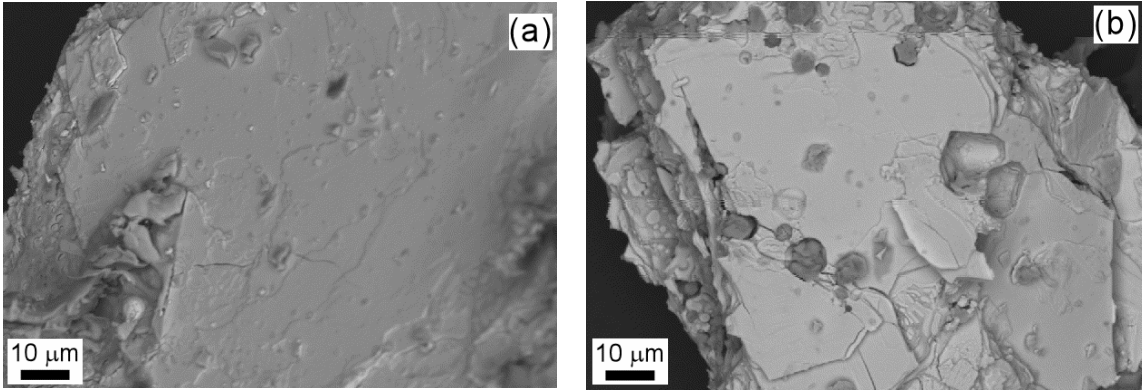

**Figure 1**. Secondary electron images of plate-like crystals for (a) *TM*-free and (b) Co-co-doped (Ca,La)112 bulk samples.

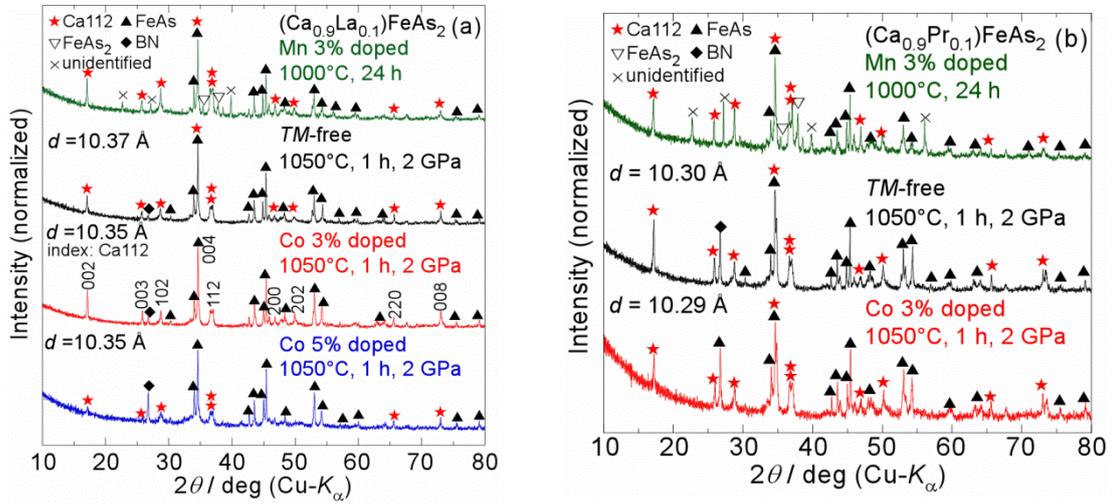

**Figure 2**. Powder XRD patterns of *TM*-free, Mn or Co-co-doped (a) (Ca,La)112 and (b) (Ca,Pr)112 samples sintered at 1000°C for 24 h in evacuated quartz ampoules, or at 1050°C for 1 h under 2 GPa.

Analyzed compositions of *RE* and Co in high pressure synthesized samples by EDX were summarized in Table 1. Analyzed *RE* compositions were larger than the nominal composition of *RE* in all samples. However, these values are smaller than that of a $(Ca_{0.85}La_{0.15})FeAs_2$ sample reported by Sala *et al*. where La/(La+Ca) = 0.25[12]. On the other hand, actual compositions of Co were close to the nominal ones.

**Table 1**. Analyzed compositions of *RE* and Co in (Ca,*RE*)112 synthesized under high pressure.

| nominal composition | $RE/(RE+Ca)$ | $Co/(Fe+Co)$ |
|---|---|---|
| $(Ca_{0.9}La_{0.1})FeAs_2$ | 0.170 | - |
| $(Ca_{0.9}La_{0.1})Fe_{0.97}Co_{0.03}As_2$ | 0.149 | 0.026 |
| $(Ca_{0.9}La_{0.1})Fe_{0.95}Co_{0.05}As_2$ | 0.132 | 0.047 |
| $(Ca_{0.9}Pr_{0.1})FeAs_2$ | 0.170 | - |
| $(Ca_{0.9}Pr_{0.1})Fe_{0.97}Co_{0.03}As_2$ | 0.203 | 0.036 |

Figures 3(a) and 3(b) show the temperature dependences of the zero-field-cooled (ZFC) and field-cooled (FC) magnetization curves of *TM*-free and co-doped (Ca,*RE*)FeAs$_2$ samples, respectively. *TM*-free (Ca,La)112 and (Ca,Pr)112 samples showed superconductivity with $T_c$ of 38 and 23 K, respectively. These values are higher than $T_c$ of the samples synthesized from nominal compositions of $(Ca_{0.85}RE_{0.15})FeAs_2$ ($T_c$ = 24.5 K(*RE*=La), 13.2 K(Pr))[12]. The tendency is consistent with the report by Kudo *et al.*[11] in which an increase in La concentration at Ca-site above 15% decreases $T_c$ of (Ca,La)112. Mn-co-doping suppressed superconductivity of both (Ca,La)112 and (Ca,Pr)112. On the other hand, improvement of superconducting properties in Co-co-doped sample was observed. Co-co-doped (Ca,Pr)112 sample showed higher $T_c$ = 36 K than *TM*-free (Ca,Pr)112, and Co-co-doped (Ca,La)112 showed sharper transition than *TM*-free sample though $T_c$ was not changed largely (38 K(*TM*-free), 38 K(Co 3% co-doped) and 36 K(Co 5% co-doped)). Furthermore, Co-co-doped (Ca,La)112 showed large diamagnetic screening below $T_c$, suggesting bulk superconductivity. At present the reason for relatively high $T_c$ by direct doping to Fe site is unclear, but EDX analysis showed that La composition in (Ca,La)112 decreased in Co-co-doped (Ca,La)112, similar to decrease of La composition in Sb-co-doped (Ca,La)112[13]. Optimization of As-Fe-As bond angle might be another reason of improvement of superconducting properties as observed in Sb-co-doped samples[13].

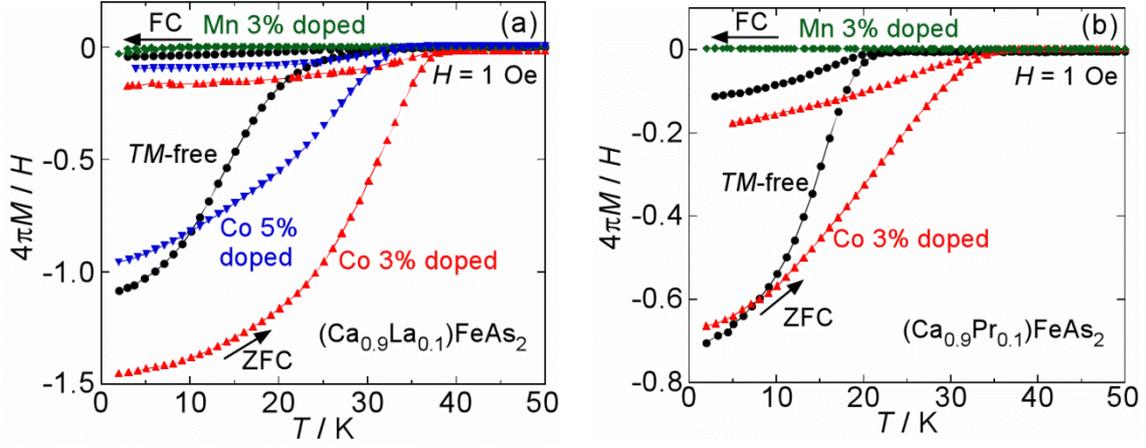

**Figure 3**. Temperature dependence of the ZFC and FC magnetization curves of *TM*-free, Mn or Co-co-doped (a) (Ca,La)112 and (b) (Ca,Pr)112 samples sintered at 1000°C for 24 h in evacuated quartz ampoules, or 1050°C for 1 h under 2 GPa.

Magnetic hysteresis loops measured at 2, 5, 10, 20, 25 K and $J_c$ are shown in Figs. 4(a) and 4(b), respectively. $J_c$ was calculated from the width of magnetization hysteresis based on the extended Bean model

$$J_c = \frac{20\Delta M}{a(1 - a/3b)}$$

where $\Delta M$(emu cm$^{-3}$) = $M^+$ - $M^-$ ($M^-$ and $M^+$ are magnetizations measured with field increasing and decreasing, respectively) and $a$(cm) and $b$(cm) are lengths of shorter and longer edges of the sample. Co-co-doped (Ca,La)112 showed wider magnetization hysteresis than *TM*-free (Ca,La)112, and the $J_c$ values near zero field were estimated to be 2.1 x 10$^4$ Acm$^{-2}$ and 3.2 x 10$^3$ Acm$^{-2}$ at 2 K and 25 K, respectively, though *TM*-free (Ca,La)112 have low $J_c$ (2.5 x 10 Acm$^{-2}$ at 2 K near zero field). These relatively high $J_c$ values as for a polycrystalline bulk also indicate the bulk superconductivity of (Ca,*RE*)112 at least below 25 K. The $J_c$ under magnetic field is relatively high (5.2 x 10$^3$ Acm$^{-2}$ at 2 K, 4.8 T), suggesting that (Ca,La)112 have intrinsically high $J_c$ under magnetic fields similar to other iron based superconductors. It also suggests the improvement of grain boundaries in Co-co-doped (Ca,La)112.

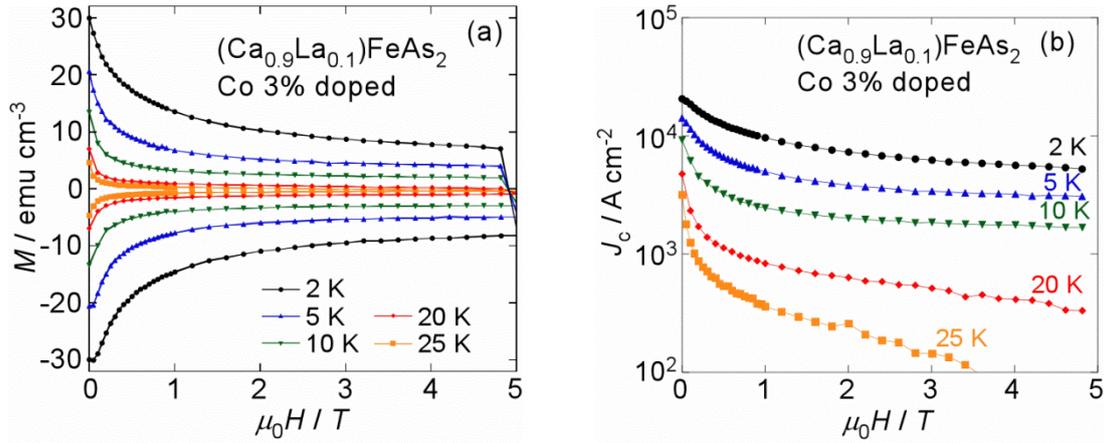

**Figure 4**. (a) Magnetic hysteresis loops of a Co(3%) co-doped (Ca,La)112 sample measured at 2-25 K under $H // c$.

(b) $J_c$-$H$ curves for a Co(3%) co-doped (Ca,La)112 sample calculated from the extended Bean model.

Figures 5(a) and 5(b) show temperature dependence of resistivity of the *TM*-free and Co-co-doped (Ca,La)112 sintered bulks, respectively, synthesized at 1050°C for 1 h under 2 GPa. Resistivity curves measured under various magnetic fields are shown in insets. Spikes observed in r-T curves in high temperature region (>200 K) is possibly due to unstabe attachment between probes ans samples. The resistivity of *TM*-free and Co-co-doped (Ca,La)112 at 300 K was 13.8 and 24.0 mΩ, respectively. These values are much higher than that of single crystal of (Ca,La)112 (0.26 mΩ at 300 K)[9]. The high resistivity in sintered bulk samples indicates that samples have weak grain connections and the measured resistivity was not intrinsic. $T_c^{onset}$ were 42 and 39 K under 0 T in *TM*-free and Co-co-doped samples, respectively. However, superconducting transition of *TM*-free sample was very broad, and $T_c^{zero}$ of *TM*-free sample was 14 K. On the other hand, $T_c^{zero}$ of Co-co-doped sample was 30 K. This sharp transition also suggests that Co-co-doping improved grain boundaries in the (Ca,La)112 bulk sample.

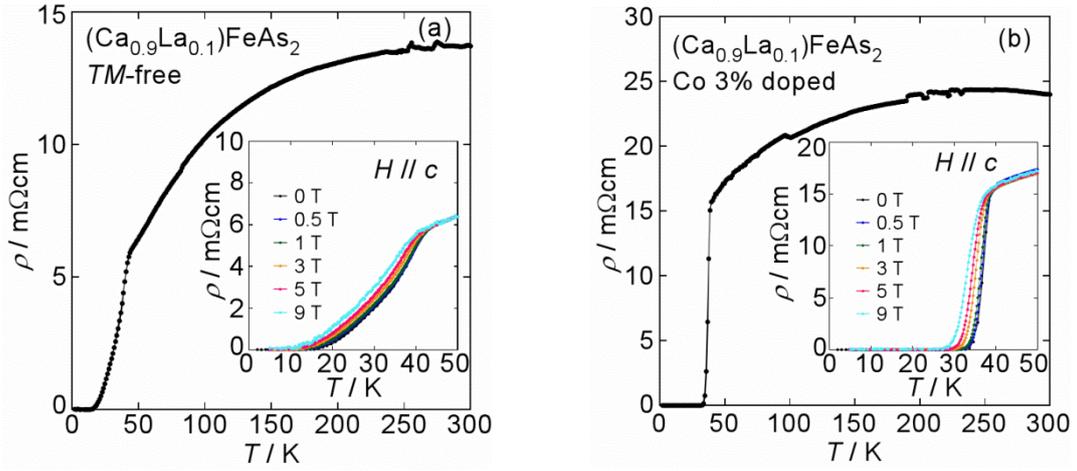

**Figure 5**. Temperature dependence of the resistivity of (a) *TM*-free and (b) Co-co-doped (Ca,La)112 polycrystalline bulk samples measured under various magnetic fields applied parallel to the *c*-axis.

Figure 6 shows temperature dependence of $H_{c2}$ of *TM*-free and Co-co-doped (Ca,La)112 determined by 90% and 50% of the normal state resistivity ($\rho_N$) at $T_c^{onset}$ = 42 and 39 K for *TM*-free and Co-co-doped sample, respectively. $H_{c2}(0)$ values estimated from Werthamer-Helfand-Hohenberf (WHH) model[19] using 90 % $\rho_N$ were ~92.5 T(d $\mu_0 H$/d$T$ = -3.33 T K$^{-1}$) and ~92.5 T(d $\mu_0 H$/d$T$ = -3.51 T K$^{-1}$) in *TM*-free and Co-co-doped (Ca,La)112, respectively. When we use 50 % $\rho_N$ for determination of $H_{c2}(0)$, ~54.7 T(d $\mu_0 H$/d$T$ = -2.35 T K$^{-1}$) and ~61.6 T(d $\mu_0 H$/d$T$ = -2.41 T K$^{-1}$) were obtained for *TM*-free and Co-co-doped (Ca,La)112, respectively. These $H_{c2}(0)$ values under $H$ // $c$ were higher than 39.4 T ($H$ // $c$) reported for a *TM*-free (Ca$_{0.77}$La$_{0.18}$)112 single crystal[14].

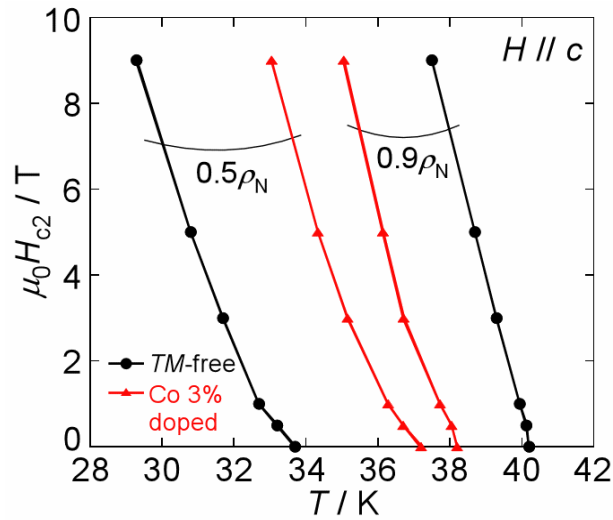

**Figure 6**. Temperature dependence of $H_{c2}$ for *TM*-free and Co(3%)-doped (Ca$_{0.9}$La$_{0.1}$)112 determined by 90% and 50% of the normal state resistivity($\rho_N$).

## 4. Summary

In summary, *TM*-co-doping effect on superconducting properties of (Ca,*RE*)112 (*RE* = La, Pr) was investigated. Mn-co-doping for (Ca,*RE*)112 suppressed superconductivity, while Co-co-doped (Ca$_{0.9}$Pr$_{0.1}$)112 showed higher $T_c$ and Co-co-doped (Ca$_{0.9}$La$_{0.1}$)112 showed sharper transition than *TM*-free samples. These results suggest that Co-co-doping improved As-Fe-As bond angles in (Ca,*RE*)112 samples. La concentration in (Ca,La)112 was decreased by Co-co-doping, this also contributed to the sharp transition probably. If optimal doping level of *RE* and transition metals is established, further increase of $T_c$ can be expected. In the resistivity measurement, large enhancement of $T_c^{zero}$ suggesting improvement of grain boundaries by Co-co-doping was observed. Critical current density of La and Co-co-doped sample was approximately 2.1 x 10$^4$ Acm$^{-2}$ at 2 K, 3.2 x 10$^3$ Acm$^{-2}$ at 25 K near zero field, respectively. These relatively high $J_c$ values indicate bulk superconductivity of (Ca,*RE*)112.

## 5. Acknowledgements

This work was partly supported by the JSPS KAKENHI Grant Number 26390045, Izumi Science and Technology Foundation, and European Union-Japan project SUPER-IRON (grant agreement No. 283204).

**Reference**

[1] Kamihara Y, Watanabe T, Hirano M and Hosono H 2008 *J. Am. Chem. Soc*. **130** 3296-7

[2] Hsu F C, Luo J Y, Yeh K W, Chen T K, Huangh T W, Wu P M, Lee Y C, Huang Y L, Chu Y Y, Yan D C and Wu M K 2008 *Proc. Natl. Acad. Sci. U.S.A*. **23** 14262-4

[3] Pitcher M J, Parker D R, Adamson P, Herkelrath S J C, Boothroyd A T, Ibberson R M, Bruneli M and Clarke S J 2008 *Chem. Commun*. 5918-20

[4] Rotter M, Tegel M and Johrendt D 2008 *Phys. Rev. Lett*. **101** 107006

[5] Ogino H, Matsumura Y, Katsura Y, Ushiyama K, Horii S, Kishio K and Shimoyama J 2009 *Supercond. Sci. Technol*. **22** 075008

[6] Kakiya S, Kudo K, Nishikubo Y, Oku K, Nishibori E, Sawa H, Yamamoto T, Nozaka T and Nohara M 2011 *J. Phys. Soc. Jpn*. **80** 093704

[7] Sun Y L, Jianf H, Zhai H F, Bao J K, Jiao W H, Tao Q, Shen C Y, Zeng Y W, Xu Z A and Cao G H 2012 *J. Am. Chem. Soc*. **134** 12893-6

[8] Katrych S, Rogacki K, Pisoni A, Bosma S, Weyeneth S, Gaal R, Zhigadlo N D, Karpinski J and Forró L 2013 *Phys. Rev. B* **87** 180508

[9] Katayama N, Kudo K, Onari S, Mizukami T, Sugawara K, Sugiyama Y, Kitahama Y, Iba K, Fujimura K, Nishimoto N, Nohara M and Sawa H 2013 *J. Phys. Soc. Jpn*. **82** 123702

[10] Yakita H, Ogino H, Okada T, Yamamoto A, Kishio K, Tohei T, Ikuhara Y, Gotoh Y, Fujihisa H, Kataoka K, Eisaki H and Shimoyama J 2014 *J. Am. Chem. Soc*. **136** 846-9

[11] Kudo K, Mizukami T, Kitahama Y, Mitsuoka D, Iba K, Fujimura K, Nishimoto N, Hiraoka Y and Nohara M 2014 *J. Phys. Soc. Jpn*. **83** 025001

[12] Sala A, Yakita H, Ogino H, Okada T, Yamamoto A, Kishio K, Ishida S, Iyo A, Eisaki H, Fujioka M, Takano Y, Putti M and Shimoyama J 2014 *Appl. Phys. Express* **7** 073102

[13] Kudo K, Kitahama Y, Fujimura K, Mizukami T, Ota H and Nohara M 2014 *J. Phys. Soc. Jpn*. **83** 093705

[14] Zhou W, Zhuang J, Yuan F, Li X, Xing X, Sun Y and Shi Z 2014 *Appl. Phys. Express* **7** 063102

[15] Sefat A S, Jin R, McGuire M A, Sales B C, Singh D J and Mandrus D 2008 *Phys. Rev. Lett*. **101** 117004

[16] Sefat A S, Huq A, McGuire M A, Jin R, Sales B C, Mandrus D, Cranswick L M D, Stephens P W and Stone K H 2008 *Phys. Rev. B* **78** 104505

[17] Wang C, Li Y K, Zhu Z W, Jiang S, Lin X, Luo Y K, Chi S, Li L J, Ren Z, He M, Chen H, Wang Y T, Tao Q, Cao G H and Xu Z A 2009 *Phys. Rev. B* **79** 054521

[18] Chang P, Xiang Z J, Ye G J, Lu X F, Lei B, Wang A F, Chen F and Luo X G 2014 *Supercond. Sci. Technol.* 27 065012

[19] Werthamer N R, Helfand E and Hohenberg P C 1966 *Phys. Rev.* **147** 295-302